\documentclass [12 pt] {article}

\usepackage{latexsym}
\usepackage{amsmath}
\usepackage{amsfonts}
\usepackage[dvips]{color}

\allowdisplaybreaks

\def\d{\mbox{\rm d}}

\def\ie{{\it ie }}

\def\pa#1#2{\displaystyle{\frac{\partial#1}{\partial#2}}}

\def\dddot#1{\mathinner{\buildrel\vbox{\kern5pt\hbox{...}}\over{#1}}}
\def\ddddot#1{\mathinner{\buildrel\vbox{\kern5pt\hbox{....}}\over{#1}}}

\begin {document}

\begin{center}
{\huge On Differential Sequences}\\[4 mm]
\large{K Andriopoulos\footnote{Author to whom correspondence should be addressed.}, PGL Leach and A Maharaj\\[2 mm] School of Mathematical Sciences, Westville Campus,\\[2mm] University of KwaZulu-Natal, Durban 4000,\\[2mm] Republic of South Africa}\footnotetext[2]{\makeatletter Emails: kand@aegean.gr; leachp@ukzn.ac.za; ADHIRMAHARAJ@telkomsa.net \makeatother}
\\[7 mm]
\end {center}
\linespread{1.3}
\abstract {We introduce the notion of Differential Sequences of ordinary differential equations. This is motivated by related studies based on evolution partial differential equations. We discuss the Riccati Sequence in terms of symmetry analysis, singularity analysis and identification of the complete symmetry group for each member of the Sequence. The singularity analysis reveals considerable structure for the values of the coefficients of the leading-order terms and resonances of the different principal branches. Full proofs of the symmetry properties are performed for differential equations defined by their recursion properties and not given in explicit form.}

\strut\hfill

\section{Introduction}

\strut\hfill

In the study and application of differential equations there are certain equations which are prominent by means of their utility of application or their ubiquity of occurrence.  In the particular case of ordinary differential equations examples of members of this prominent class are the Riccati equation, the linear second-order equation, the linear third-order equation of maximal symmetry, the Ermakov-Pinney equation, the Kummer-Schwarz equation and its generalisation.  These equations are not independent.  As Conte observed \cite {Conte 94 a}, the study of any of the Riccati equation, the linear second-order equation and the Kummer-Schwartz equation is equivalent to a study of the other two.  More recently the closely connected linear third-order of maximal symmetry and the Ermakov-Pinney equation have been added to the list given by Conte.  In \cite {Euler 06 b} the generalised Kummer-Schwartz equation has also been included.  Since these equations are of different orders and have differing symmetry properties, it is evident that the connections among them are nonlocal.

\strut\hfill

A considerable part of the motivation for this study is found in the works of Peterssen {\it et al} \cite {Petersen 04 a} and Euler {\it et al} \cite {Euler 03 a} in which the authors obtain recursion operators for linearisable $1 +1 $ evolution equations.  The specific equation of relevance to this work is the eighth of their classification, {\it videlicet}
\[
u_t=u_{xx}+\lambda_8u_x+h_8u_x^2,
\]
where $\lambda_8$ is an arbitrary constant and $h_8$ is an arbitrary function of the dependent variable $u$.  The corresponding recursion operator is
\[
\mathbf{VIII} \qquad R_8[u]=D_{x}+h_8u_{x}.
\]

The usual area of application of recursion operators has been that of partial differential equations and in particular evolution equations of which the equation above is a sample.  Euler {\it et al} \cite{Euler 06 a} applied the idea of a recursion operator to ordinary differential equations.  They did not present a treatment of all possible classes, but concentrated upon two representative equations, {\it videlicet} the Riccati equation and the Ermakov-Pinney equation.

\strut\hfill

The recursion operator which generates the Riccati Sequence\footnote{In \cite{Euler 06 a} the term `Riccati Hierarchy' is used. For a reason which becomes apparent in what follows we believe that the term `sequence' is more appropriate. The `correct' term to be used to describe this sequence has been a matter of perception of the language.} is
\begin{equation}
\mbox{\rm R.O.} = D + y, \label{RO}
\end{equation}
where $D$ denotes total differentiation with respect to $x$.  The recursion operator
essentially originates from Case VIII, $R_8[u]$, for partial differential equations when the dependence
on $t$ is removed, {\it ie} $(u_t = 0)$, and one writes $u_x = y$.
We note that in applications to ordinary differential equations the notation usually used in the context of partial differential equations can be and is simplified.

\strut\hfill

When (\ref{RO}) acts on\footnote{This is a matter of notation. In this draft we have chosen the simplest. In another context, especially when one wishes to be `physically' correct, one would choose $\exp [\int y\d x] $ since, if $y$ is replaced by $-x$ one obtains the generating function for the solution of the time-independent Schr\"odinger equation for the simple harmonic oscillator.} $y$ (which can be denoted as $R_0$) it generates, in succession, all the members of the Sequence which we denote by $R_n$, {\it videlicet}
\begin{eqnarray}
y' + y^2 &=& 0 \quad (R_1) \label{R1} \\
y'' + 3 yy' + y^3 &=& 0 \quad (R_2) \label{R2} \\
y''' + 4yy'' + 3y'{}^2 + 6y^2y' + y^4 &=& 0 \quad (R_3) \label{R3} \\
y'''' + 5yy''' + 10y'y'' + 10y^2y'' + 15yy'{}^2 + 10y^3y' + y^5 &=& 0 \quad (R_4) \label{R4} \\
&\vdots& \nonumber\\
(D + y)^n y &=& 0 \quad (R_n). \label{Rn}
\end{eqnarray}
As an aid in our discussion we distinguish between an element of the Sequence, denoted as indicated above by $R_n $, and its left side by writing the latter as $\tilde {R}_n $.

\strut\hfill

In addition to the formal definition given in (\ref {Rn}) there is a differential recurrence relation.

\strut\hfill

\noindent {\bf Lemma\label{Lemma 1}:} {\it The members of the Riccati Sequence satisfy the differential recurrence relation}
\begin {equation}
\frac {\partial} {\partial y}\left (\tilde {R}_{m+ 1}\right) = (m+ 2)\tilde {R}_m. \label {1.1}
\end {equation}

\strut\hfill

\noindent {\bf Proof:}  By inspection of (\ref {R1}), (\ref {R2}) and (\ref {R3}) it is evident that (\ref {1.1}) is true for the initial members of the sequence.  We assume that
\begin {equation}
\frac {\partial} {\partial y}\left (\tilde {R}_{k+ 1}\right) = (k+ 2)\tilde {R}_k \label {1.2}
\end {equation}
for some $k \geq {2} $.  Then
\begin {eqnarray*}
\displaystyle {\frac {\partial} {\partial y}\left (\tilde {R}_{k+ 2}\right)} & = & \displaystyle {\frac {\partial } {\partial y}\left [(D + y)\tilde {R}_{k+ 1}\right]} \\
& = &\left [(D + y)\partial_y+ 1\right]\tilde {R}_{k+ 1} \\
& = & (D + y) (k+ 2)\tilde {R}_k+ \tilde {R}_{k+ 1} \\
& = & (k+ 3)\tilde {R}_{k+ 1}
\end {eqnarray*}
as required.

\noindent {\bf Remark:}  This recurrence relation is a little intriguing bearing in mind the connection to the generating function for the Hermite polynomials.  One recalls the differential recurrence relation for the Hermite polynomials, {\it videlicet}
$$
\frac {\d H_n (x)} {\d x} = 2nH_{n-1} (x),
$$
and notes the similarity of structure to that given in (\ref {1.1}).

\strut\hfill

The Riccati Sequence contains as its two lower members the Riccati equation and the Painlev\'e-Ince equation.  The Riccati equation \cite {Riccati 24 a} has an history now of almost three centuries and is one of the few examples known for which there exists a nonlinear superposition principle.  Its close relationship to the linear second-order differential equation and the Kummer-Schwartz equation has already been noted \cite{Conte 94 a}.  The Painlev\'e-Ince equation \cite {Painleve 02 a, Ince 27 a} is a notable example of a nonlinear second-order differential equation of maximal symmetry possessing both Left and Right Painlev\'e Series \cite {Feix 97 a} and arises in a remarkable number of applications\footnote {For a listing see \cite {Mahomed 85 a, Feix 97 a, Chandrasekar 07 a, Andriopoulos 06 b}.  The authors of the third paper place (\ref {R2}) in a broader class of equations which they describe as being of modified Emden type.}.

\strut\hfill

In this paper we report the notable properties of the members of the Riccati Sequence.  We concentrate upon the number of Lie point symmetries, singularity properties, first integrals, explicit integrability and complete symmetry groups.  In Section 2 we present the symmetry analysis of the members of the Sequence.  In terms of the explicit integrability of the members of the Sequence the lack of point symmetry in the higher-order members is remarkable and indicates the necessity for considering nonlocal symmetries.  We note that this was the case in the treatment of a pair of equations of Ermakov-Pinney type \cite {Leach 05 a}.  In Section 3 we elaborate upon the results of the singularity analysis of the lower members of the Riccati Sequence presented by Euler {\it et al} \cite {Euler 06 a} by considering the pattern of the values of the resonances for the general member of the Sequence.  Section 4 is devoted to a consideration of the invariants and first integrals of the members of the Sequence. A short discussion on the symmetries of the integrals is presented. In Section 5 we present the complete symmetry group of the general member of the Sequence after firstly considering the results for $R_4$ to give a concrete basis for the theoretical discussion.  In Section 6 we recall the solution of $R_n $ and extend our discussion to an equation containing combinations of the $R_i $.  In the case of the second-order equation so formed the properties have been known for a long time \cite {Painleve 02 a, Ince 27 a, Kamke 83 a}.  Among our concluding remarks in Section 6 we summarise the remarkable properties found for the Riccati Sequence which is based upon a rather elementary recursion operator.  We indicate that the route to complexification presented in Section 6 broadens the class of differential equations which may be subsumed into a more general concept of a Riccati Sequence. 

\strut\hfill

\section{Symmetry analysis}

\strut\hfill

As a first-order ordinary differential equation (\ref{R1}) possesses an infinite number of Lie point symmetries. Equation (\ref{R2}) was examined in \cite{Mahomed 85 a} for its Lie point symmetries which were found to be the following eight, indeed the maximal number for a second-order ordinary differential equation, here written
in a more elegant way to match the results of the succeeding members of the sequence,
\begin{eqnarray*}
\Gamma_1 &=& x^2y\partial_x - y[2+xy(xy - 2)]\partial_y  \\
\Gamma_2 &=& y\partial_x - y^3\partial_y \\
\Gamma_3 &=& xy\partial_x + y^2(1 - xy)\partial_y \\
\Gamma_4 &=& x\partial_x- y\partial_y \\
\Gamma_5 &=& x^3(xy-2)\partial_x - x(xy -2)[2+xy(xy - 2)]\partial_y \\
\Gamma_6 &=& -x^2(xy -2)\partial_x + xy(xy -2)(xy-1)\partial_y \\
\Gamma_7 &=& x ^2\partial_x+ (2- 2xy)\partial_y \\
\Gamma_8 &=& \partial_x
\end{eqnarray*}
with the algebra $sl(3,R)$.

\strut\hfill

We proceed with $R_3$ to find an unexpected three, {\it ie}
\begin{eqnarray*}
\Gamma_1 &=& \partial_x\\
\Gamma_2 &=& x\partial_x - y\partial_y\\
\Gamma_3 &=& x^2\partial_x + (3-2xy)\partial_y
\end{eqnarray*}
with the algebra $sl(2,R)$.

\strut\hfill

{\bf Proposition:} {\it The general member of the Riccati Sequence possesses the symmetries}
\begin{eqnarray*}
\Gamma_1 &=& \partial_x\\
\Gamma_2 &=& x\partial_x - y\partial_y\\
\Gamma_3 &=& x^2\partial_x + (n-2xy)\partial_y
\end{eqnarray*}
{\it with the algebra} $sl(2,R)$.

{\bf Remark:} {For $R_2$ there are an additional five symmetries
and for $R_1$ an additional infinity.}

\strut\hfill

{\bf Proof:} The possession of $\Gamma_1$ is straightforward since (\ref{Rn}) is autonomous. We proceed with $\Gamma_2$.
The $n $th extension of $\Gamma_2 $ is
\begin {equation}
\Gamma_2 ^ {[n]} = x\partial_x- \sum_{j = 0} ^n (j+ 1)y ^ {(j)}\partial_{y ^ {(j)}} \label {20.1}
\end {equation}
and it is a simple calculation to demonstrate that
\begin {eqnarray*}
\Gamma_2^ {[1]}\tilde {R}_1 & = & -2\tilde {R}_1 \\
\Gamma_2^ {[2]}\tilde {R}_2 & = & -3\tilde {R}_2.
\end {eqnarray*}
We assume that the statement
\begin {equation}
\Gamma_2^ {[m]}\tilde {R}_m  =  - (m+ 1)\tilde {R}_m \label {20.2}
\end {equation}
is correct and we prove that $$\Gamma_2^ {[m+1]}\tilde {R}_{m+1}  =  - (m+ 2)\tilde {R}_{m+1}.$$
We commence with the statement
\begin {equation}
\Gamma_2^ {[m+ 1]}\tilde {R}_{m+ 1}  =  \Gamma_2^ {[m+ 1]} (D +y)\tilde {R}_m. \label {20.3}
\end {equation}
We break the calculation of the right side of (\ref {20.3}) into some parts.  Firstly we have
\begin {equation}
\Gamma_2 ^ {[m+ 1]}D =\left [\Gamma_2 ^ {[m+ 1]}, D\right]_{LB} + D\Gamma_2 ^ {[m+ 1]}. \label {20.4}
\end {equation}
The Lie Bracket is
\begin {eqnarray}
& &\left [x\partial_x-\sum_{j = 0} ^ {m+ 1} (j+ 1)y ^ {(j)}\partial_{y ^ {(j)}}, \partial_x+ \sum_{k = 0} ^ {\cdot}y ^ {(k+ 1)}\partial_{y ^ {(k)}}\right]_{LB}  \hspace*{70mm}   \nonumber\\
& & \hspace*{20mm} = -\partial_x- \left [\sum_{j = 0} ^ {m+ 1} (j+ 1)y ^ {(j)}\partial_{y ^ {(j)}}, \sum_{k = 0} ^ {\cdot}y ^ {(k+ 1)}\partial_{y ^ {(k)}}\right]_{LB}     \nonumber\\
& & \hspace*{20mm} = -\partial_x- \sum_{j = 0} ^ {m+ 1} (j+ 1)\sum_{k = 0} ^ {\cdot}\left [y ^ {(j)}\partial_{y ^ {(j)}}, y ^ {(k+ 1)}\partial_{y ^ {(k)}}\right]_{LB}     \nonumber\\
& & \hspace*{20mm} = -\partial_x- \sum_{j = 0} ^ {m+ 1} (j+ 1)\sum_{k = 0} ^ {\cdot}\left\{y ^ {(j)}\delta_{j,k+ 1}\partial_{y ^ {(k)}} - y ^ {(k+ 1)}\delta_{k,j}\partial_{y ^ {(j)}}\right\} \nonumber \\
& & \hspace*{20mm} = -\partial_x- \sum_{j = 0} ^ {m+ 1} (j+ 1)\left\{y ^ {(j)}\partial_{y ^ {(j-1)}} - y ^ {(j+ 1)}\partial_{y^ {(j)}}\right\} \nonumber \\
& & \hspace*{20mm} = -\partial_x- \sum_{j = 0} ^ {m} (j+ 2)y ^ {(j+ 1)}\partial_{y ^ {(j)}} +  \sum_{j = 0} ^ {m+ 1} (j+ 1)y ^ {(j+ 1)}\partial_{y ^ {(j)}} \nonumber \\
& & \hspace*{20mm} = -\partial_x- \sum_{j = 0} ^ {m} y ^ {(j+ 1)}\partial_{y ^ {(j)}} + (m+ 2)y ^ {(m+ 2)}\partial_{y ^ {(m+ 1)}} \nonumber \\
& & \hspace*{20mm} = - D + (m+ 2)y ^ {(m+ 2)}\partial_{y ^ {(m+ 1)}}, \nonumber 
\end {eqnarray}
where the overdot on the summation means that the sum is taken to whatever order of derivative required.

\strut\hfill

In a much simpler calculation we have
\begin {eqnarray}
\Gamma_2^ {[m+ 1]} y\tilde {R}_m & = &\left (\Gamma_2 ^ {[m+ 1]} y\right)\tilde {R}_m+ y\left (\Gamma_2^ {[m+ 1]}\tilde {R}_m\right) \nonumber \\
& = & - y\tilde {R}_m+ y\Gamma_2^ {[m+ 1]}\tilde {R}_m. \label {20.6}
\end {eqnarray}
We substitute the results of both the Lie Bracket and (\ref {20.6}) into (\ref {20.3}).  We have
\begin {eqnarray}
\Gamma_2^ {[m+ 1]}\tilde {R}_{m+ 1} & = &  \left\{  - D + (m+ 2)y ^ {(m+ 2)}\partial_{y ^ {(m+ 1)}}  + D\Gamma_2 ^ {[m+ 1]}\right. \nonumber \\
& & \left.  - y+ y\Gamma_2^ {[m+ 1]}\right\}\tilde {R}_m \nonumber \\
& = & - (D +y)\tilde {R}_m+ (D +y)\Gamma_2^ {[m+ 1]}\tilde {R}_m \nonumber \\
& \stackrel{(\ref{20.2})}{=} & - (D +y)\tilde {R}_m- (D +y) (m+ 1)\tilde {R}_m \nonumber \\
& = & - (m+ 2) (D +y)\tilde {R}_m \nonumber \\
& = & - (m+ 2)\tilde {R}_{m+ 1} \label {20.7}
\end {eqnarray}
which proves the result.

\strut\hfill

Note that in the proof we have made use of the fact that $\tilde {R}_m $ contains derivatives only up to $y ^ {(m)} $.

\strut\hfill

We turn now to the proof in the case of $\Gamma_3 $.

\strut\hfill

It is a simple matter to show that
\begin {eqnarray*}
\Gamma_3^ {[1]}\tilde {R}_1 & = & -2x (1+ 1)\tilde {R}_1 \\
\Gamma_3^ {[2]}\tilde {R}_2 & = & -2x (2+ 1)\tilde {R}_2\\
\Gamma_3^ {[3]}\tilde {R}_3 & = & -2x (3 + 1)\tilde {R}_3
\end {eqnarray*}
and one can easily assume that
\begin {equation}
\Gamma_3^ {[m]}\tilde {R}_m =  -2x (m + 1)\tilde {R}_m \label {20.8}
\end {equation}
so that the task is now to demonstrate that from this property it follows that
\begin {equation}
\Gamma_3^ {[m+ 1]}\tilde {R}_{m+ 1}  =  -2x (m + 2)\tilde {R}_{m+ 1}. \label {20.9}
\end {equation}

\strut\hfill

We commence with the definition and so we have
\begin {eqnarray*}
\Gamma_3^ {[m+ 1]}\tilde {R}_{m+ 1}  & = & \Gamma_3^ {[m+ 1]} (D +y) \tilde {R}_{m}  \\
& = & \left [\Gamma_3^ {[m+ 1]}, (D +y)\right]_{LB} \tilde {R}_{m} + (D +y)  \Gamma_3^ {[m+ 1]} \tilde {R}_{m} \\
& = & \left [\Gamma_3^ {[m+ 1]}, (D +y)\right]_{LB} \tilde {R}_{m} + (D +y) \left ( \Gamma_3^ {[m]} +\partial_y\right) \tilde {R}_{m},
\end {eqnarray*}
where the last line is a consequence of remembering the definition of $\Gamma_3$ and that the $(m+1)th$ derivative in $\Gamma_3^ {[m+1]}$ does not act on  $\tilde {R}_{m}$.

\strut\hfill

The first task is to compute the Lie Bracket. We have
\begin {eqnarray*}
& &\left [x ^2\partial_x+ (m+ 1 -2xy)\partial_y-2x\sum_{j = 1} ^ {m+ 1} (j+ 1)y ^ {(j)}\partial_{y ^ {(j)}} -\sum_{j = 1} ^ {m+ 1} j (j+ 1)y ^ {(j-1)}\partial_{y ^ {(j)}}, \right. \hspace{50mm}\\
& & \hspace{7mm} \left. \partial_x+\sum_{k = 0} ^ {\cdot }y ^ {(k+ 1)}\partial_{y ^ {(k)}} +y \right] \\
&  & \hspace{13mm}= \,\,  -2x\partial_x+2y\partial_y+2\sum_{j = 1} ^ {m+ 1} (j+ 1)y ^ {(j)}\partial_{y ^ {(j)}}  +2xy'\partial_y+ (m+ 1 -2xy) \\
&  & \hspace{20mm} +2x\sum_{j = 1} ^ {m+ 1} (j+ 1)\sum_{k = 0} ^ {\cdot }\left [y ^ {(k+ 1)}\partial_{y ^ {(k)}} ,\,y ^ {(j)}\partial_{y ^ {(j)}} \right]_{LB} \\
&  & \hspace{20mm} +\sum_{j = 1} ^ {m+ 1} j (j+ 1)\sum_{k = 0} ^ {\cdot }\left [y ^ {(k+ 1)}\partial_{y ^ {(k)}} ,y ^ {(j-1)}\partial_{y ^ {(j)}} \right]_{LB}.
\end {eqnarray*}
We compute the two subsidiary Lie Brackets separately.
\begin {eqnarray*}
\mbox {\rm LB}_1 & = & 2x\sum_{j = 1} ^ {m+ 1} (j+ 1)\sum_{k = 0} ^ {\cdot }\left \{y ^ {(k+ 1)}\delta_{k,j}\partial_{y ^ {(j)}} - y ^ {(j)}\delta_{j,k+ 1}\partial_{y ^ {(k)}}\right\} \\
& = & 2x\sum_{j = 1} ^ {m+ 1} (j+ 1)\left \{y ^ {(j+ 1)}\partial_{y ^ {(j)}} - y ^ {(j)}\partial_{y ^ {(j-1)}}\right\} \\
& = & 2x\left\{(m+ 2)y ^ {(m+ 2)}\partial_{y ^ {(m+ 1)}} - 2y'\partial_y - \sum_{j = 1} ^ {m} y ^ {(j+ 1)}\partial_{y ^ {(j)}} \right\}.
\end {eqnarray*}
In a similar way we find that
\[
\mbox {\rm LB}_2  = (m+ 1) (m+ 2)y ^ {(m+ 1)}\partial_{y ^ {(m+ 1)}} - 2y\partial_y - 2\sum_{j = 1} ^ {m} (j+ 1)y ^ {(j)}\partial_{y ^ {(j)}}.
\]
We return to the main calculation and after some simplification of terms which cancel we continue as below.
\begin {eqnarray*}
\Gamma_3^ {[m+ 1]}\tilde {R}_{m+ 1}  & = & \left\{-2x\left (\partial_x+y'\partial_y + \sum_{j = 1} ^ {m} y ^ {(j+1)}\partial_{y ^ {(j)}} \right) + 2 (m+ 2)y ^ {(m+ 1)}\partial_{y ^ {(m+ 1)}} \right. \\
&  &\left. +m+ 1 -2xy + (m+ 1) (m+ 2)y ^ {(m+ 1)}\partial_{y ^ {(m+ 1)}}  \right. \\
&  &\left.  + 2x (m+ 2)y ^ {(m+ 2)}\partial_{y ^ {(m+ 1)}} + (D +y)\left (-2x (m + 1) +\partial_y\right)\right\}\tilde {R}_{m} \\
& = & \left\{-2x(D +y) + (m+1) + (D +y)\partial_y - 2(m+ 1)(D+y) x\right\}\tilde {R}_m \\
& = & -2x (m+ 2) (D +y)\tilde {R}_m+\left [(D +y)\partial_y - (m+ 1)\right]\tilde {R}_m \\
& = & -2x (m+ 2)\tilde {R}_{m+ 1}
\end {eqnarray*}
since
$$
\left [(D +y)\partial_y- (m+ 1)\right]\tilde {R}_m = 0.
$$
This last statement is equivalent to the result of Lemma \ref {Lemma 1}, {\it ie}
$$
\frac {\partial} {\partial y}\left (\tilde {R}_{m+ 1}\right) = (m+ 2)\tilde {R}_m
$$
and so the result is proven.

\noindent {Remark:}  Note that in the proof above we have again made use of the fact that $\tilde {R}_m $ contains derivatives only up to $y ^ {(m)} $.

\strut\hfill

\section{Singularity analysis}

\strut\hfill

We use singularity analysis as a tool to determine whether a given differential equation is integrable in terms of functions almost everywhere analytic. For equations which pass the Painlev\'{e} Test we are then encouraged to seek closed-form solutions. We find that all members of the Riccati Sequence possess the Painlev\'{e} Property. As we see below, explicit integrability follows. Independently of this integrability the results of the singularity analysis provide some very interesting patterns in terms of the parameters, that is the $p$, $\alpha$ and $r$ of the standard analysis. Similarly interesting patterns have been reported in \cite{Leach 07 a}.

\strut\hfill

Euler {\it et al} \cite{Euler 06 a} presented the singularity analysis of the Riccati Sequence. For all elements the leading-order behaviour is $\alpha\chi^{-1}$ with the possible values of $\alpha$ being listed in the Tables. We summarise their results. 

\strut\hfill

\begin{center}
Table I be here
\end{center}

\strut\hfill

In Table II we advance from \cite{Euler 06 a} and present the properties in terms of the singularity analysis for the general member of the Riccati Sequence. In that way we are able to comment upon the integrability or otherwise of all members of the Sequence in the sense of Painlev\'e.

\strut\hfill

\begin{center}
Table II be here
\end{center}

\strut\hfill

The pattern of the resonances is as follows: The set of resonances for $\alpha = j$ is obtained from the set
for $\alpha = j-1$ by subtracting the number $n+1$ from the largest positive resonance of the latter set.

\strut\hfill

It follows from Tables I and II that all members of the Riccati Sequence pass the Painlev\'e Test and therefore
each is integrable in terms of analytic functions. 

\strut\hfill

When we apply the Riccati transformation
\begin{equation}
y = \alpha \frac{w'}{w} \label{RT1}
\end{equation}
on the $n$th member of the Riccati Sequence, we observe that the most simplified equation, {\it ie} $w^{(n+1)} = 0$, arises when we chose $\alpha = 1$ which is a consequence of the singularity analysis itself \cite{Andriopoulos 06a}.
Therefore (\ref{RT1}) may be written as
$$
x = x,\quad w = \exp\left[\int y\d x\right].
$$
It is a matter of simple calculation to verify the following proposition.

\strut\hfill

{\bf Proposition:} {\it  The general solution of the $n$th member of the Riccati Sequence, $n\geq 1$, is given by
\begin{equation}
y_n = \frac{\left(\sum_{i=0}^{n} A_i x^i\right)'}{\sum_{i=0}^{n} A_i x^i}, \label{2.22a}
\end{equation}
where the $A_i$, $i=0,n$, are constants of integration.}

\strut\hfill

\section{Invariants and First Integrals}

\strut\hfill

For the purposes of this paper we commence with some definitions.

\strut\hfill

{\bf Definition 1:} An invariant $I$ of an ordinary differential equation is said to be any nontrivial
function which satisfies
\begin{equation}
\frac{\d I}{\d x} = 0 \label{def}
\end{equation}
on solution curves of the differential equation.

\strut\hfill

{\bf Definition 2:} A first integral of an $n$th-order ordinary differential equation is any function $I = I(x, y, y', \ldots, y^{(n-1)})$ which satisfies (\ref{def}).

\strut\hfill

Note that we do depart from some standard definitions for reasons which become evident below.  There are various conventions concerning the meanings of the expressions `first integral', `invariant' and `conserved quantity'.  We are not concerned by the third in this paper since we do not use that expression.  Some writers use the first two expressions interchangeably.  Others prefer to distinguish between the two by insisting that the former be autonomous whereas the latter is allowed to depend upon the independent variable.  This is a sensible distinction under appropriate circumstances.  Indeed in terms of the integration of an ordinary differential equation the distinction can be quite critical.  In this paper we have varied the definitions to suit the very precise purpose of distinguishing between two classes of function both of which have the property of having a zero total derivative on solution curves of the differential equation.

\strut\hfill

We wish to identify all invariants and first integrals, as defined
above, of each member of the Riccati Sequence. In order to do so
we take the $n$th member of that Sequence, perform an increase of
order by using the Riccati transformation to obtain $w^{(n+1)} =
0$. We compute the fundamental integrals and invariants of that
equation and by reverting the transformation one can deduce the whole set of invariants for the $n$th member of the Riccati Sequence, {\it videlicet}
\begin{equation}
I_j = \left(\sum_{i=1}^j \frac{(-1)^{i+1}}{(j-i)!}x^{j-i}\tilde{R}_{(n-i)}\right)\exp\left[\int y\d x\right], \quad j= 1, n+1,
\label{inv}
\end{equation}
where $\tilde{R}_{-1} = 1$ by convention.\newline
Note that for the invariants of $w^{(n)} = 0$ we would have the above invariants (\ref{inv}) except that instead of
$\tilde{R}_{(n-i)}$ we would have $w^{(n+1-i)}$ and there would be no exponential.

\strut\hfill

We now turn our attention to first integrals. One takes
the ratio of two separate invariants to obtain a first integral.
An independent set for $R_n $ can be variously defined.  The set $\{{\cal F}_{ij}\}$, defined by
\begin {equation}
{\cal F}_{ij} = \frac {I_j} {I_i},\quad j = 1,i- 1,i+ 1,n, \label
{30.1}
\end {equation}
is an independent set of first integrals for $R_n $.  Such a simple formula is not available if one wishes to describe a set of autonomous first integrals.

\strut\hfill

An interesting aspect arises in the symmetry properties of the first integrals which we briefly note. 
If one computes the symmetries of all first integrals of (\ref{R2}), one finds that they all share the algrebra $A_1\oplus_s A_2$. By a curious misfortune this feature does not persist and therefore one is confronted with an absolute zero for contact (not to mention point) symmetries for the first integrals of (\ref{R3})\footnote{When equation (\ref{R3}) is related to $w^{(iv)} = 0$ by an increase of order, the symmetries of the first integrals of the latter become, by inverting the transformation, nonlocal symmetries for the ones of the former.}.

\strut\hfill

\section{Complete symmetry groups}

\strut\hfill

The concept of a complete symmetry group of a differential equation was introduced by
Krause \cite{Krause 94 a} as the group associated with the set of
symmetries, be they point, contact, generalised or nonlocal,
required to specify the equation or system completely.  

\strut\hfill

We start with $R_4$ to give a flavor of the procedure and then we prove the general result for any member
of the Riccati Sequence.

\strut\hfill

{\bf Proposition:} {\it  The complete symmetry group of $R_4$ is given by the symmetries}
\begin{eqnarray*}
\Delta_1 &=& - \exp\left[-\int y \d x\right]\{y\}\partial_y \\
\Delta_2 &=& - \exp\left[-\int y \d x\right]\{xy - 1\}\partial_y \\
\Delta_3 &=& - \exp\left[-\int y \d x\right]\{x^2y - 2x\}\partial_y \\
\Delta_4 &=& - \exp\left[-\int y \d x\right]\{x^3y - 3x^2\}\partial_y \\
\Delta_5 &=& - \exp\left[-\int y \d x\right]\{x^4y - 4x^3\}\partial_y.
\end{eqnarray*}

\strut\hfill

{\bf Proof:} We calculate the fourth extensions of $\Delta_1 - \Delta_5$.
\begin{eqnarray*}
\Delta_1^{[4]} &=& - \exp\left[-\int y \d x\right]\{y\partial_y + (y' - y^2)\partial_{y'} + (y'' - 3yy' + y^3)\partial_{y''} \\
&& + (y''' - 4yy'' - 3y'^2 + 6y^2y' - y^4)\partial_{y'''} \\
&& + (y^{(iv)} - 5yy''' - 10y'y'' + 10y^2y'' + 15yy'{}^2 - 10y^3y' + y^5)\partial_{y^{(iv)}}\} \\
\Delta_2^{[4]} &=& - \exp\left[-\int y \d x\right]\{-\partial_y + 2y\partial_{y'} + 3(y'-y^2)\partial_{y''} + 4(y'' - 3yy' +y^3)\partial_{y'''} \\
&&+ 5(y''' - 4yy'' - 3y'^2 + 6y^2y' - y^4)\partial_{y^{(iv)}}\} + x\Delta_1^{[4]} \\
\Delta_3^{[4]} &=& - 2\exp\left[-\int y \d x\right]\{-\partial_{y'} + 3y\partial_{y''} + 6(y'-y^2)\partial_{y'''} + 10(y'' - 3yy' + y^3)\partial_{y^{(iv)}}\} \\
&& + 2x\Delta_{2eff}^{[4]} + x^2\Delta_{1}^{[4]} \\
\Delta_4^{[4]} &=& - 6\exp\left[-\int y \d x\right]\{-\partial_{y''} + 4y\partial_{y'''} + 10(y'-y^2)\partial_{y^{(iv)}}\} \\
&& + 3x\Delta_{3eff}^{[4]} + 3x^2\Delta_{2eff}^{[4]} + x^3\Delta_1^{[4]} \\
\Delta_5^{[4]} &=& - 24\exp\left[-\int y \d x\right]\{-\partial_{y'''} + 5y\partial_{y^{(iv)}}\} \\
&& + 4x\Delta_{4eff}^{[4]} + 6x^2\Delta_{3eff}^{[4]} + 4x^3\Delta_{2eff}^{[4]} + x^4\Delta_1^{[4]},
\end{eqnarray*}
where the subscript {\it eff} stands for the effective part of that symmetry.\newline
When we act all the above extensions on the general fourth-order ordinary differential equation, {\it videlicet}
$$
y^{(iv)} = f(x,y,y',y'',y'''),
$$
we obtain the system of five equations
\begin{eqnarray}
y^{(iv)} - 5yy''' - 10y'y'' + 10y^2y'' + 15yy'{}^2 - 10y^3y' + y^5 = y\pa{f}{y} + (y' -y^2)\pa{f}{y'}&& \nonumber\\ + (y'' -3yy' + y^3)\pa{f}{y''} + (y''' - 4yy'' - 3y'^2 + 6y^2y' - y^4)\pa{f}{y'''}&& \label{45} \\
&&\nonumber\\
5(y''' - 4yy'' - 3y'^2 + 6y^2y' - y^4) = -\pa{f}{y} + 2y\pa{f}{y'} && \nonumber\\ + 3(y'-y^2)\pa{f}{y''} + 4(y'' -3yy' + y^3)\pa{f}{y'''} && \nonumber\\ &&\nonumber\\
10(y'' - 3yy' + y^3) = -\pa{f}{y'} + 3y\pa{f}{y''} + 6(y'-y^2)\pa{f}{y'''} && \nonumber\\ &&\nonumber\\
10(y'-y^2) = -\pa{f}{y''} + 4y\pa{f}{y'''}&& \nonumber \\ &&\nonumber\\
5y = -\pa{f}{y'''},&& \nonumber
\end{eqnarray}
which can be solved backwardsly to give the following expressions for the derivatives of all arguments in $f$
\begin{eqnarray*}
\pa{f}{y'''} &=& -5y \\
\pa{f}{y''} &=& -10(y'+y^2) \\
\pa{f}{y'} &=& -10(y'' +3yy' +y^3) \\
\pa{f}{y} &=& -5(y''' + 4yy'' + 3y'^2 + 6y^2y' + y^4).
\end{eqnarray*}
When these expressions are substituted into (\ref{45}) we recover $R_4$.

\strut\hfill

{\bf Proposition:} {\it  The complete symmetry group of $R_n$ is given by the $(n+1)$ symmetries\footnote{The correct number of symmetries required to specify an equation completely has been discussed in \cite{Andriopoulos 01} and \cite{Andriopoulos 02}.}}
$$
\Delta_i = - \exp\left[-\int y \d x\right]\{x^{i-1}y - (i-1)x^{i-2}\}\partial_y, \,\, i=1, n+1.
$$

{\bf Proof:} We remind the reader that $R_n$ stands for the $n$th member of the Riccati Sequence and $\tilde{R}_n$
for the left hand side of $R_n$. It is essential for the notation required in this proof to introduce
the adjoint recursion operator, $R.O.^{\alpha} = D - y$, which generates the adjoint Riccati Sequence, {\it videlicet}
\begin{eqnarray}
y' - y^2 &=& 0 \quad (R^{\alpha}_1) \label{Ra1} \\
y'' - 3 yy' + y^3 &=& 0 \quad (R^{\alpha}_2) \label{Ra2} \\
y''' - 4yy'' - 3y'{}^2 + 6y^2y' - y^4 &=& 0 \quad (R^{\alpha}_3) \label{Ra3} \\
y^{(iv)} - 5yy''' - 10y'y'' + 10y^2y'' + 15yy'{}^2 - 10y^3y' + y^5 &=& 0 \quad (R^{\alpha}_4) \label{Ra4} \\
y^{(v)} -6yy^{(iv)} - 15y'y''' + 15y^2y''' + 60yy'y'' - 10y''^2 - 20y^3y'' && \nonumber\\ + 15y'^3 - 45y^2y'^2 + 15y^4y' -  y^6 &=& 0 \quad (R^{\alpha}_5) \label{Ra5} \\
&\vdots& \nonumber\\
(D - y)^n y &=& 0 \quad (R^{\alpha}_n). \label{Ran}
\end{eqnarray}
The left hand side of the above members is denoted by $\tilde{R}^{\alpha}_n$. 

\strut\hfill

\noindent {\bf Remark:} We note that there is a reflection here between the Riccati Sequence and its adjoint in that the formul{\ae} for the Sequence are reflected in the formul{\ae} for the adjoint Sequence.

\strut\hfill

\noindent {\bf Lemma\label{Lemma 2}:} {\it The general members of the Riccati and the adjoint Riccati Sequences satisfy the recurrence relation}
\begin {equation}
\tilde{R}_{n} = \tilde{R}_n^\alpha + \sum_{i=1}^n {n+1 \choose i} \tilde{R}_{i-1}^\alpha \tilde{R}_{n-i}. \label{55}
\end {equation}

\strut\hfill

\noindent {\bf Proof:} It is easy to show that (\ref{55}) holds for $n=2$ and $n=3$. We assume that (\ref{55}) is true for $n=k$ and we prove that it is true for $n=k+1$.
\begin{eqnarray*}
\tilde{R}_{k+1} &=& (D + y) \tilde{R}_{k} = (D+ y) \left[\tilde{R}_{k}^\alpha + \sum_{i=1}^k {k+1 \choose i} \tilde{R}_{i-1}^\alpha \tilde{R}_{k-i}\right] \\
&=& \tilde{R}_{k+1}^\alpha + 2y\tilde{R}_{k}^\alpha + \sum_{i=1}^k {k+1 \choose i} \left\{D\left[\tilde{R}_{i-1}^\alpha \tilde{R}_{k-i}\right] + y\tilde{R}_{i-1}^\alpha \tilde{R}_{k-i}\right\} \\
&=& \tilde{R}_{k+1}^\alpha + 2y\tilde{R}_{k}^\alpha + \sum_{i=1}^k {k+1 \choose i} \left(\tilde{R}_{i}^\alpha + y\tilde{R}_{i-1}^\alpha\right)\tilde{R}_{k-i}\\ && \quad + \sum_{i=1}^k {k+1 \choose i} \tilde{R}_{i-1}^\alpha\left(\tilde{R}_{k-i+1} - y\tilde{R}_{k-i}\right) + y\sum_{i=1}^k {k+1 \choose i}\tilde{R}_{i-1}^\alpha\tilde{R}_{k-i} \\
&\stackrel{(n=k)}{=}& \tilde{R}_{k+1}^\alpha + 2y\tilde{R}_{k}^\alpha + \sum_{i=1}^k {k+1 \choose i}\tilde{R}_{i}^\alpha\tilde{R}_{k-i} + \sum_{i=1}^k {k+1 \choose i} \tilde{R}_{i-1}^\alpha\tilde{R}_{k-i+1} + y(\tilde{R}_{k} - \tilde{R}_{k}^\alpha) \\
&=& \tilde{R}_{k+1}^\alpha + y\tilde{R}_{k}^\alpha + y\tilde{R}_{k} + \sum_{i=2}^{k+1} {k+1 \choose i-1}\tilde{R}_{i-1}^\alpha\tilde{R}_{k-i+1} + \sum_{i=1}^k {k+1 \choose i} \tilde{R}_{i-1}^\alpha\tilde{R}_{k-i+1} \\
&=& \tilde{R}_{k+1}^\alpha + y\tilde{R}_{k}^\alpha + y\tilde{R}_{k} + \sum_{i=1}^{k+1}\left\{ {k+1 \choose i-1} + {k+1 \choose i}\right\} \tilde{R}_{i-1}^\alpha\tilde{R}_{k-i+1} - \tilde{R}_{k}^\alpha\tilde{R}_{0} -\tilde{R}_{0}^\alpha\tilde{R}_{k} \\
&=& \tilde{R}_{k+1}^\alpha + \sum_{i=1}^{k+1}{k+2 \choose i} \tilde{R}_{i-1}^\alpha\tilde{R}_{k-i+1},
\end{eqnarray*}
which proves the result.

\strut\hfill

The $n$th extension of $\Delta_i$ is
\begin{eqnarray*}
\Delta_i^{[n]} &=& - \exp\left[-\int y \d x\right]\left\{-(i-1)!\partial_{y^{(i-2)}} + \frac{i!}{1!}y\partial_{y^{(i-1)}} + \frac{(i+1)!}{2!}\tilde{R}^{\alpha}_1\partial_{y^{(i)}} \right.\\ 
&& \left. + \frac{(i+2)!}{3!}\tilde{R}^{\alpha}_2\partial_{y^{(i+1)}} +\ldots + \frac{(n+1)!}{(n-i+2)!}\tilde{R}^{\alpha}_{n-i+1}\partial_{y^{(n)}}\right\} \\
&& + \sum_{k=0}^{i-2} \frac{1}{k!}\frac{\d ^k}{\d x^k}(x^{i-1})\Delta_{(k+1)eff}^{[n]}, \quad i = 1,n+1,
\end{eqnarray*}
where
$$ \Delta_{(1)eff}^{[n]} = \Delta_{1}^{[n]}, \partial_{y^{(-1)}} = 0, \,\, \partial_{y^{(0)}} = \partial_{y}, \,\, \partial_{y^{(1)}} = \partial_{y'}, \,\, {\rm etc}, \,\, \frac{\d ^0}{\d x^0} = 1, \,\, \frac{\d ^1}{\d x^1} = \frac{\d}{\d x} \,\, {\rm etc}. $$
The action of the effective part of $\Delta_i^{[n]}$,
$$
\Delta_{(i)eff}^{[n]} = -\partial_{y^{(i-2)}} + \sum_{j=0}^{n-i+1} {i+j \choose j+1}\tilde{R}^{\alpha}_j\partial_{y^{(i-1+j)}},
$$
on the general $n$th-order ordinary differential equation, {\it videlicet}
$$
y^{(n)} = f(x,y,y',y'',y''',\ldots,y^{(n-1)}),
$$
gives
\begin{equation}
{n+1 \choose n-i+2}\tilde{R}^{\alpha}_{n-i+1} = -\displaystyle{\frac{\partial f}{\partial y^{(i-2)}}} + \sum_{j=0}^{n-i} {i+j \choose j+1}\tilde{R}^{\alpha}_j\displaystyle{\frac{\partial f}{\partial y^{(i-1+j)}}}, \,\, i=1,n+1. \label{star}
\end{equation}
For $i=1$ (\ref{star}) gives
\begin{equation}
\tilde{R}^{\alpha}_{n} = \sum_{j=0}^{n-1} \tilde{R}^{\alpha}_{j}\frac{\partial f}{\partial y^{(j)}}. \label{130}
\end{equation}
The remaining $n$ of (\ref{star}) are equivalent to
\begin{equation}
{n+1 \choose n-i+1}\tilde{R}^{\alpha}_{n-i} = -\displaystyle{\frac{\partial f}{\partial y^{(i-1)}}} + \sum_{j=i+1}^{n} {j \choose j-i}\tilde{R}^{\alpha}_{j-i-1}\displaystyle{\frac{\partial f}{\partial y^{(j-1)}}}, \,\, i=1,n. \label{startwo}
\end{equation}
System (\ref{startwo}) may be conveniently represented as the system of equations
\begin{equation}
{\bf L}_n^\alpha = {\bf Q}_n^\alpha {\bf F}_n, \label{p.1}
\end{equation}
where
\begin{eqnarray}
({\bf L}_n^\alpha)_i &=& {n+1 \choose i} \tilde{R}_{n-i}^\alpha, \qquad i=1,n, \nonumber\\
({\bf F}_n)_i &=& \pa{f}{y^{(i-1)}}, \quad i=1,n, \nonumber\\
({\bf Q}_n^\alpha)_{ij} &=& \left\{\begin{array}{rl} {\displaystyle{{j \choose i}}} \tilde{R}_{j-i-1}^\alpha, & i<j,\\ -1, & i=j,\\ 0, & i>j.\end{array}\right. \label{p.105}  
\end{eqnarray}
We define
\begin{equation}
({\bf Q}_n)_{ij} = \left\{\begin{array}{rl} -{\displaystyle{{j \choose i}}} \tilde{R}_{j-i-1}, & i<j,\\ -1, & i=j,\\ 0, & i>j.\end{array}\right. \label{p.205}
\end{equation}
and
$$
({\bf L}_n)_i = {n+1 \choose i} \tilde{R}_{n-i}, \qquad i=1,n.
$$

We prove the following two Lemmas.

\strut\hfill

\noindent {\bf Lemma\label{Lemma 3}:} {\it The matrices ${\bf Q}_n^\alpha$ and ${\bf Q}_n$ defined in (\ref{p.105}) and (\ref{p.205}) are the inverses of each other, \ie they satisfy}  
\begin {equation}
{\bf Q}_n^\alpha {\bf Q}_n = {\bf I}_n. \label{p.4}
\end {equation}

\strut\hfill

\noindent {\bf Proof:} That (\ref{p.4}) is true for $n=1$ is obvious. We assume that it is true for $n=k$, \ie
\begin{equation}
{\bf Q}_k^\alpha {\bf Q}_k = {\bf I}_k. \label{p.5}
\end{equation}
Then 
\begin{eqnarray*}
{\bf Q}_{k+1}^\alpha {\bf Q}_{k+1} &=& \left[\begin{array}{rr} {\bf Q}_k^\alpha & {\bf L}_k^\alpha \\ {\bf O}_k^T & -1 \end{array}\right] \left[\begin{array}{rr} {\bf Q}_k & -{\bf L}_k \\ {\bf O}_k^T & -1 \end{array}\right] \\
&=& \left[\begin{array}{rr} {\bf Q}_k^\alpha {\bf Q}_k, & -{\bf Q}_k^\alpha {\bf L}_k - {\bf L}_k^\alpha \\ {\bf O}_k^T & 1 \end{array}\right] \\ &\stackrel{(\ref{p.5})}{=} & {\bf I}_{k+1}
\end{eqnarray*}
provided ${\bf Q}_k^\alpha {\bf L}_k + {\bf L}_k^\alpha = 0$. We consider the $i$th element of this term.
\begin{eqnarray*}
({\bf Q}_k^\alpha {\bf L}_k + {\bf L}_k^\alpha)_i &=& \sum_{j=1}^k ({\bf Q}_k^\alpha)_{ij} ({\bf L}_k)_j + ({\bf L}_k^\alpha)_i \\ &=& -({\bf L}_k)_i + \sum_{j=i+1}^k ({\bf Q}_k^\alpha)_{ij} ({\bf L}_k)_j + ({\bf L}_k^\alpha)_i \\ &=& - {k+1 \choose i} \tilde{R}_{k-i} + \sum_{j=i+1}^k {j \choose i} \tilde{R}_{j-i-1}^\alpha {k+1 \choose j} \tilde{R}_{k-j} + {k+1 \choose i} \tilde{R}_{k-i}^\alpha \\ &=& -{k+1 \choose i} \left\{ \tilde{R}_{k-i} - \tilde{R}_{k-i}^\alpha - \sum_{j=i+1}^k {k+1-i \choose j-i} \tilde{R}_{j-i-1}^\alpha \tilde{R}_{k-j}\right\} \\ &\stackrel{(\ref{55})}{=}& 0.
\end{eqnarray*}

\strut\hfill

\noindent {\bf Lemma\label{Lemma 4}:} 
\begin{equation}
{\bf Q}_n {\bf L}_n^\alpha = - {\bf L}_n. \label{lemmafour} 
\end{equation}

\strut\hfill

\noindent {\bf Proof:} Consider an element of the product ${\bf Q}_n {\bf L}_n^\alpha$. This is
\begin{eqnarray*}
({\bf Q}_n)_{ij}({\bf L}_n^\alpha)_j &=& -{n+1 \choose i} \tilde{R}_{n-i}^\alpha + \sum_{j=i+1}^n \left[-{j \choose i} \tilde{R}_{j-i-1}\right] \\ && \qquad \times {n+1 \choose j}\tilde{R}_{n-j}^\alpha \\
&=& - {n+1 \choose i} \left\{\tilde{R}_{n-i}^\alpha + \sum_{j=i+1}^n {j \choose i}{n+1 \choose j}/{n+1 \choose i}\right. \\ && \left.\phantom{{n+1 \choose i}} \times \tilde{R}_{n-j}^\alpha \tilde{R}_{j-i-1}\right\} \\
&\stackrel{(\ref{55})}{=}& - {n+1\choose i}\tilde{R}_{n-i} \\
&=& - ({\bf L}_n)_i.
\end{eqnarray*}

\strut\hfill

Therefore (\ref{p.1}) is written as
\begin{eqnarray}
{\bf F}_n &=& \left({\bf Q}_n^\alpha\right)^{-1} {\bf L}_n^\alpha \nonumber\\ 
&\stackrel{(\ref{p.4})}{=}& {\bf Q}_n {\bf L}_n^\alpha \nonumber\\
&\stackrel{(\ref{lemmafour})}{=}& - {\bf L}_n. \label{p.2}
\end{eqnarray}
Equation (\ref{130}) is equally written as
$$
\tilde{R}^{\alpha}_{n} = \sum_{i=1}^{n} \tilde{R}^{\alpha}_{i-1}\frac{\partial f}{\partial y^{(i-1)}}, 
$$
which, when (\ref{p.2}) and (\ref{55}) are used, results to $(R_n)$ and the proposition is proven.

\strut\hfill

\section{Concluding remarks}

\strut\hfill

We are familiar with the use of sequences of numbers and functions in the mathematical and wider scientific literature. The last several decades have seen intensive study of hierarchies based upon nonlinear evolution equations which have their bases in the mathematical modelling of physical phenomena.  The extension to ordinary differential equations has only been made recently \cite {Euler 06 a}.  We have chosen to term these related equations as differential sequences to reflect the dual nature of a definition which combines the idea of an operation which generates the elements of the sequence and also that the elements of the sequence are composed of derivatives.  It seems to us that the word sequence is more appropriate in this context than hierarchy and is more in keeping with the tradition of mathematical terminology.

\strut\hfill

The Riccati Sequence, which has been the subject of the present study, illustrates with a degree of excellence, which one hopes can be surpassed, the type of mathematical properties which are likely to make such sequences an object of fond study.  The Riccati equation has a long and distinguished history in both the theory of differential equations and the application to divers phenomena.  The Riccati Sequence and its adjoint are based on recursion operators closely related in form to the Dirac operators of the quantum mechanical simple harmonic oscillator which in themselves are simply autonomous versions of two of the Lie point symmetries of the classical simple harmonic oscillator.  With such a lavish heritage one should not be too surprised that the Riccati Sequence\footnote {Equally the adjoint sequence.  In the following discussion the properties of the adjoint sequence may be inferred {\it mutatis mutandis} from those of the Riccati Sequence.}  exhibits such properties in terms of symmetry and integrability.

\strut\hfill

Each element of the sequence can be linearised by means of the nonlocal transformation -- often called a Riccati transformation -- and so is trivially integrable.  Although $R_1 $ and $R_2 $ -- the Riccati and Painlev\'e-Ince equations -- display exceptional symmetry in the sense of Lie point symmetries, the remaining elements of the sequence possess just the three-element algebra $sl (2,R) $ of Lie point symmetries.  The distinguishing feature of the Lie symmetries of the elements of the sequence is the possession of $n+ 1 $ (in the case of $R_n $) exponential nonlocal symmetries which completely specify the elements.  The algebra of these symmetries is $(n+ 1) A_1 $.

\strut\hfill

As a consequence of the ability to linearise the equations of the Riccati Sequence each element possesses an invariant derived from the so-called fundamental integrals of the parent linear equation\footnote {These integrals have been termed fundamental since an $n $th-order linear ordinary differential equation possesses $n $ linearly independent integrals which are linear in the dependent variable and its first $(n- 1) $ derivatives.  The integrals are known for their interesting algebraic properties.}.  These invariants are not integrals in the conventional sense as they contain the integral of the dependent variable.  However, the integral of the dependent variable appears as an exponential term and so functions of these invariants which are homogeneous of degree zero in the dependent variable are first integrals.  It was for this reason that we introduced specialised meanings for the two expressions, `invariant' and `first integral', for the purposes of this paper.  Since there are $(n+ 1) $ linearly independent (exponential nonlocal) invariants, there are $n $ functionally independent integrals which reflect most adequately the integrability of each member of the Riccati Sequence.

\strut\hfill

In terms of singularity analysis the Riccati Sequence possesses properties which may even be regarded to outshine the symmetry properties.  Not only is each element of the Sequence explicitly integrable in terms of analytic functions apart from isolated polelike singularities -- a property which has been used to illustrate certain subtle and not widely appreciated implications of singularity analysis \cite {Andriopoulos 06 b} -- but also there are patterns to the possible values of the coefficients of the leading-order terms and the values of the resonances for each of the principal branches.  In passing one recalls that each branch is principal and the need to be concerned with the implications of branches with unfortunate properties is obviated.  The patterns of the resonances deserve further, separate treatment, particularly in respect of other possible sequences.

\strut\hfill

The Riccati Sequence has provided an excellent vehicle for the introduction of the notion of differential sequences of ordinary differential equations.  Its Recursion Operator is simple.  Its generating function is elementary.  Two of the elements, $R_1 $ and $R_2$, are well-known in the literature as well as in applications.  Already Euler {\it et al} \cite {Euler 06 a} have made a brief excursion into the properties of the Ermakov-Pinney Sequence, as the name suggests, based upon the Ermakov-Pinney equation \cite {Ermakov 80 a, Pinney 50 a} which has been found in many varied applications.

\strut\hfill

In the Introduction we have recalled the classical association of the Riccati equation, the linear second-order ordinary differential equation and the third-order Kummer-Schwarz equation.  Since these equations are related by means of nonlocal transformations, there is no reason to exclude from this select group other equations which are similarly related\footnote {One is well aware that all properties may not travel well through the process of nonlocal transformation, but that obstacle is already encountered under far less exotic transformations.}.  Already in hand are initial studies of a sequence based upon the Kummer-Schwarz equation \cite {Euler 06 b} for which the Recursion Operator is now an integrodifferential operator of third order by way of contrast with the simple Recursion Operator of the Riccati Sequence.  One looks forward to further revelations of fascinating properties of other differential sequences.

\strut\hfill

As a final remark we recall that the evolution equations and associated Recursion Operators \cite {Petersen 04 a} which are the source material for this work contain arbitrary parameters and unspecified functions.  In this work we have deliberately kept to the minimum required to make a sensible sequence.  It is a little like treating the autonomous linear oscillator rather than a time-dependent, damped and forced oscillator.  The latter can be transformed to the former by means of well-defined transformations and a study of the former is easier due to the simplicity possible in the presentation.  Despite our personal preference for simplicity we do nevertheless accept that there are those who wish to see the treatment of the more general systems.

\strut\hfill

Consider
\begin{equation}
E_n = \sum_{i=0}^n f_i(t) R_i = 0. \label{fc.1}
\end{equation}
Equation (\ref{fc.1}) is linearisable by means of the Riccati
transformation (\ref{RT1}), with $\alpha =1$, to
\begin{equation}
\sum_{i=0}^n f_i(t) w^{(i+1)} = 0. \label{fc.2}
\end{equation}
Equation (\ref{fc.2}) possesses $(n+1)$ solution symmetries and
the homogeneity symmetry.

\strut\hfill

Whilst there can be no dispute that the solution of (\ref {fc.2}) is more difficult than that of $w ^ {(n+ 1)} = 0 $, for our purposes there is no difference.  Subject to some mild conditions on the functions $f_i (t) $ -- basically that they be continuous \cite {Ince 27 a} [p 45] -- the solutions of (\ref {fc.2}) exist and that is all that is required for inclusion into the general framework of the treatment above.  An $n $th-order linear ordinary differential equation can have $n+ 1 $, $n+ 2 $ or $n+ 4 $ Lie point symmetries which represent the $n $ solution symmetries, linearity, autonomy (in the right variables) and the general $sl (2,R) $ symmetry of equations of maximal symmetry \cite {Mahomed 90 a}.  There is no gainsaying that the additional symmetries make the process of solution in closed form just that much easier.  However, all we need is the existence of the $n $ solution symmetries to provide the symmetries necessary for our purposes.  Consequently we have taken the clearer part so that the ideas and results contained in this paper be as evident as can be possible in such matters.

\strut\hfill

We look forward to a development of the subject of differential sequences of ordinary differential equations to match that of the related areas in partial differential equations.

\strut\hfill

\section*{Acknowledgements}

The work reported in this paper is part of a project funded under the Swedish-South African Agreement, Grant Number 60935.  AM is supported by bursaries from the National Research Foundation of South Africa and the University of KwaZulu-Natal.  KA thanks the State (Hellenic) Scholarship Foundation, the University of KwaZulu-Natal and the University of Patras.  PGLL thanks the University of the Aegean and the Universit\`a di Perugia for the provision of facilities while this work was being prepared and the University of KwaZulu-Natal for its continued support.

\newpage

\begin {table}
\begin{center}
\caption{Singularity analysis for the first four members of the Riccati Sequence.}
\[\begin {array}{|c|l|l|}\hline
&&\\
\mbox{\rm Member}&\mbox{\rm Leading-order coefficients}&\quad\mbox{\rm Resonances}\\
&&\\
 \hline
&&\\
R_1&\alpha = 1 &r = -1\\
&&\\
\hline
&&\\
R_2&\alpha=1 &r=-1,1\\
&\alpha = 2 & r=-1,-2 \\
&&\\
\hline
&&\\
R_3&\alpha =1 & r=-1,1,2\\
&\alpha = 2 & r=-1,1,-2\\
&\alpha =3 & r= -1,-2,-3 \\
&&\\
\hline
&&\\
R_4&\alpha = 1 & r = - 1, 1,2,3 \\
&\alpha = 2 & r=-1,1,2,-2 \\
&\alpha = 3 &r =-1,1,-2,-3 \\
&\alpha = 4 &r = -1,-2,-3,-4\\
&&\\
\hline
\end{array}\]
\end{center}
\end{table}

\newpage

\begin{table}
\begin{center}
\caption{Singularity analysis for the general member of the
Riccati Sequence.}
\[\begin {array}{|c|l|l|}\hline
&&\\
\mbox{\rm Member}&\mbox{\rm Leading-order coefficients}&\mbox{\rm Resonances}\\
&&\\
 \hline
&&\\
&   \alpha = 1 & r = -1,1,2,\ldots,n-1  \\
 R_n&  \alpha = 2 & r = - 1,1,\ldots,n- 2, - 2 \\
& \quad\,\vdots & \quad\,\vdots \\
&  \alpha = n & r = -1,-2,\ldots,-n  \\
&&\\
\hline
\end{array}\]
\end{center}
\end{table}


\begin {thebibliography} {99}

\bibitem{Andriopoulos 01}
Andriopoulos K, Leach PGL \& Flessas GP (2001) Complete symmetry groups of ordinary differential equations and their integrals: some basic considerations {\it Journal of Mathematical Analysis and Applications} {\bf 262} 256-273

\bibitem{Andriopoulos 02}
Andriopoulos K \& Leach PGL (2002) The economy of complete symmetry groups for linear higher-dimensional systems {\it Journal of Nonlinear Mathematical Physics} {\bf 9, s-2} 10-23

\bibitem {Andriopoulos 06 b}
Andriopoulos K \& Leach PGL (2006) An interpretation of the presence of both positive and negative nongeneric resonances in the singularity analysis {\it Physics Letters A} {\bf 359} 199-203

\bibitem{Andriopoulos 06a}
Andriopoulos K \& Leach PGL (2007) Autonomous self-similar ordinary differential equations and the Painlev\'{e} connection {\it Journal of Mathematical Analysis and Applications} {\bf 328} 625-639

\bibitem {Chandrasekar 07 a}
Chandrasekar VK, Senthilvelan M \& Lakshmanan M (2007) On the general solution for the modified Emden type equation $\ddot {x} +\alpha x\dot {x} +\beta x ^3= 0 $ {\it Journal of Physics A: Mathematical and Theoretical} (to appear)

\bibitem {Conte 94 a}
Conte R (1994) Singularities of differential equations and integrability in {\it Introduction to Methods of Complex Analysis and Geometry for Classical Mechanics and Nonlinear Waves} Benest D and Fr\oe schl\'e C edd (\'{E}ditions Fronti\`{e}res, Gif-sur-Yvette) 49-143

\bibitem {Ermakov 80 a}
Ermakov V (1880) Second-order differential equations.  Conditions of complete integrability {\it Universita Izvestia Kiev Series
III} {\bf 9} (1880) 1-25 (trans AO Harin)

\bibitem{Euler 03 a}
Euler M, Euler N \& Petersson N (2003) Linearizable hierarchies of evolution equations in $(1+1)$ dimensions {\it Studies in Applied Mathematics} {\bf 111} 315-337

\bibitem{Euler 06 a}
Euler M, Euler N \& Leach PGL (2006) The Riccati and Ermakov-Pinney Hierarchies {\it Journal of Nonlinear Mathematical Physics} {\bf 14} 290-302

\bibitem {Euler 06 b}
Euler M, Euler N, Leach PGL, Maharaj A \& Andriopoulos K (2007)  On generalised Kummer-Schwarz Sequences (preprint: Department of Mathematics, Lule\aa University of Technology, SE-971 87 Lule\aa, Sweden)

\bibitem{Feix 97 a}
Feix MR, G\'eronimi C, Cair\'o L, Leach PGL, Lemmer RL \& Bouquet S\'E (1997) On the singularity analysis of ordinary differential equations invariant under time translation and rescaling {\it Journal of Physics A: Mathematical and General} {\bf 30} 7437-7461

\bibitem{Ince 27 a}
Ince EL (1927) {\it Ordinary Differential Equations} (Longmans, Green \& Co, London)

\bibitem{Kamke 83 a}
Kamke E (1983) {\it Differentialgleichungen L\"osungsmethoden und L\"osungen} (BG Teubner, Stuttgart)

\bibitem {Krause 94 a}
Krause J (1994) On the complete symmetry group of the classical Kepler system {\it Journal of Mathematical Physics} {\bf 35} 5734-5748

\bibitem {Leach 05 a}
Leach PGL, Karasu (Kalkanli) A, Nucci MC \& Andriopoulos K (2005) Ermakov's superintegrable toy and nonlocal symmetries {\it SIGMA:} {\bf 1} Paper 018, 15 pages

\bibitem {Leach 07 a}
Leach PGL, Maharaj A \& Andriopoulos K (2007) Differential Sequences: The Emden - Fowler Sequence (preprint: School of Mathematical Sciences, University of KwaZulu-Natal, Durban 4041, Republic of South Africa)

\bibitem{Mahomed 85 a}
Mahomed FM \& Leach PGL (1985) The linear symmetries of a nonlinear differential equation {\it Qu\ae stiones Mathematic\ae} {\bf 8} 241-274

\bibitem {Mahomed 90 a}
Mahomed FM \& Leach PGL (1990) Symmetry Lie algebras of $n$th-order ordinary differential equations {\it Journal of Mathematical
Analysis and Applications} {\bf 151} 80-107

\bibitem{Painleve 02 a}
Painlev\'e P (1902) Sur les \'equations diff\'erentielles du second ordre et d'ordre sup\'erieur dont l'int\'egrale g\'en\'erale est uniforme {\it Acta mathematica} {\bf 25} 1-86

\bibitem {Petersen 04 a}
Petersson N, Euler N \& Euler M (2004) Recursion operators for a class of integrable third-order evolution equations {\it Studies in Applied Mathematics} {\bf 112} 201-225

\bibitem {Pinney 50 a}
Pinney E (1950) The nonlinear differential equation $y''(x)+p(x)y+cy^{-3}=0$ {\it Proceedings of the American Mathematical Society} {\bf 1} 681

\bibitem{Riccati 24 a}
Riccati Jacopo (1764) {\it Opere} (Jacopo Guisti, Lucca) 83-96

\end {thebibliography}
\end {document}